# Oblique surface waves at an interface of metal-dielectric superlattice and isotropic dielectric

**Slobodan M. Vuković[1], Juan J. Miret[2], Carlos J. Zapata-Rodriguez[3] and Zoran Jaksić[4]**

[1]Institute of Physics, University of Belgrade, Pregrevica 118, 11080 Zemun, Serbia
[2]Department of Optics, Pharmacology and Anatomy, University of Alicante, P.O. Box 99, Alicante, Spain
[3]Department of Optics, University of Valencia, Dr. Moliner 50, 46100 Burjassot, Spain
[4]Center of Microelectronic Technologies and Single Crystals, Institute of Chemistry, Technology and Metallurgy, University of Belgrade, Njegoševa 12, 11000 Belgrade, Serbia

e-mail: svukovic@ipb.ac.rs

**Abstract**. We investigate the existence and the dispersion characteristics of surface waves that propagate at an interface between metal-dielectric superlattice and isotropic dielectric. Within the long wavelength limit, when the effective-medium approximation is valid, the superlattice behaves like a uniaxial plasmonic crystal with the main optical axes perpendicular to the metal-dielectric interfaces. We demonstrate that if such a semi-infinite plasmonic crystal is cut normally to the layer interfaces and brought into the contact with semi-infinite dielectric, a new type of surface modes can appear. The propagation of such modes obliquely to the optical axes occurs under favorable conditions that regard thicknesses of the layers, as well as the proper choice of dielectric permittivity of the constituent materials. We show that losses within the metallic layers can be substantially reduced by making the layers sufficiently thin. At the same time, a dramatic enlargement of the range of angles for oblique propagation of the new surface modes is observed. This can lead, however, to the field non-locality and consequently to the failure of the effective-medium approximation.

## 1. Introduction

In contrast to the well-known surface plasmon polaritons (SPP) that can propagate along the metal-dielectric boundary, there are surface waves that exist at an interface between two transparent and homogeneous media, provided that one of them is anisotropic. Such unique type of surface waves has been termed Dyakonov surface waves (DSW) [1]. These modes are not polaritons, like SPPs, as all three components of the electric, as well as the magnetic field are involved. Thus, they are hybrid surface modes. Although DSW seem very attractive as they can propagate without losses, such modes exist under exceptionally stringent conditions. They can propagate obliquely, within a very narrow range of angles to the optical axes of anisotropic medium with positive birefringence, along the boundary with isotropic dielectric. Such conditions are very difficult to realize with natural materials. That is the main reason why it took over two decades to experimentally verify [2] theoretical predictions of Dyakonov.

As shown by Rytov back in 1955 [3], periodic media in the long wavelength limit can be considered as uniaxial crystals with optical axes perpendicular to the layers interfaces. One-dimensional photonic crystals represent such uniaxial medium. Unfortunately, periodic structures of that kind always exhibit a negative birefringence, and thus cannot be used to demonstrate the existence of DSW. In contrast to dielectric-dielectric superlattices, metal-dielectric ones (plasmonic crystals) show positive birefringence. Therefore, they are suitable to support DSW, but at the expense of introducing dispersion and losses. It is the aim of this paper to demonstrate that significant enlargement of the angular range for oblique propagation of Dyakonov-like modes can be achieved at reasonably low losses.

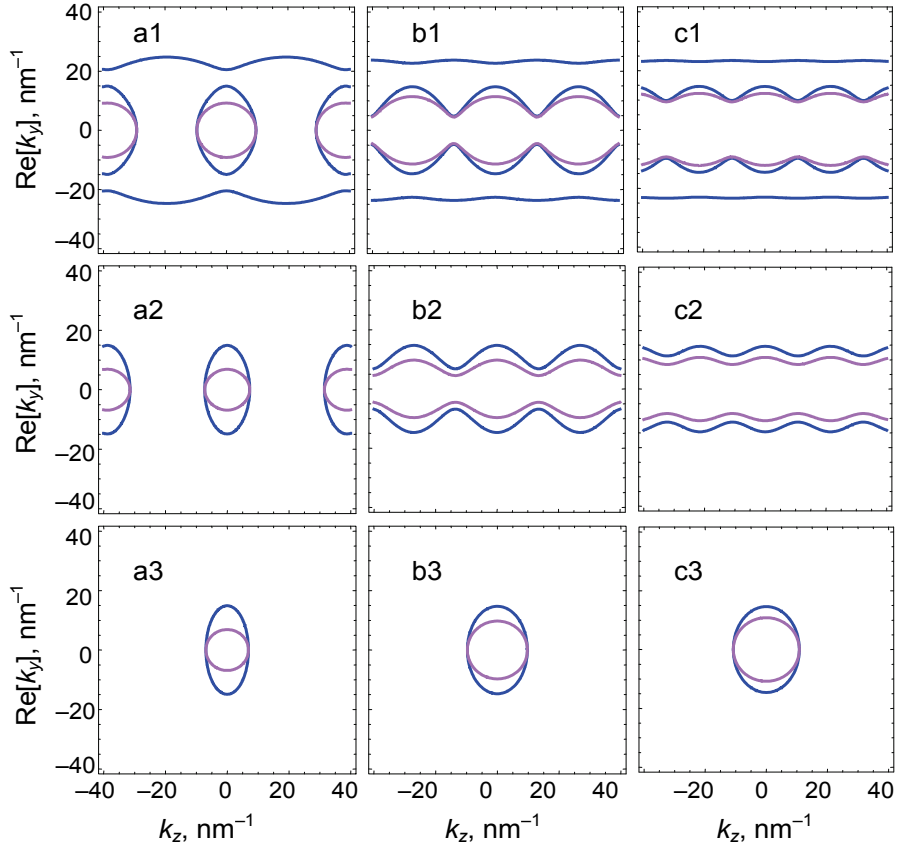

**Figure 1**. Equifrequency dispersion curves for an Ag-GaAs multilayer: $\lambda = 1.55$ μm; $\varepsilon_1 = -116$; $\varepsilon_2 = 12.4$. In all cases $d_1 = 12$ nm. 1) Exact solutions; blue – TM; magenta – TE polarization; 2) QEM approach; 3) conventional EM theory; a) $d_2 = 150$ nm; b) $d_2 = 180$ nm; c) $d_2 = 230$ nm.

## 2. Infinite metal-dielectric superlattice

We consider a nanostructured metamaterial formed as a periodic layered structure with binary metal-dielectric unit cells. We denote dielectric permittivity of metallic layers by $\varepsilon_1$, and their thickness $d_1$, while the corresponding quantities in dielectric layers are $\varepsilon_2$, and $d_2$. The size of the unit cell is $d = d_1 + d_2$. When such metamaterial is infinite, and when the wavelength of the radiation is much longer than the size of the unit cell, it is usually assumed that the Rytov [3] or effective-medium (EM) approximation is valid, and the plasmonic crystal can be represented by a diagonal permittivity tensor with elements $\{\varepsilon_\perp, \varepsilon_\perp, \varepsilon_\parallel\}$, where

$$\varepsilon_\perp = \frac{\varepsilon_1 d_1 + \varepsilon_2 d_2}{d}; \quad \varepsilon_\parallel = \frac{\varepsilon_1 \varepsilon_2 d}{\varepsilon_1 d_2 + \varepsilon_2 d_1} \quad (1)$$

The indices $\perp$, $\parallel$ indicate direction normal (x and y-directions), or parallel to the optical axes (z-direction), respectively. Then, general dispersion relation decays into two equations for the TE-polarized and TM-polarized modes, with respect to metal-dielectric interfaces

TE: $\qquad k_x^2 + k_y^2 + k_z^2 = k_0^2 \varepsilon_\perp \qquad (2)$

TM: $\qquad \dfrac{k_z^2}{\varepsilon_\perp} + \dfrac{K_y^2 + k_x^2}{\varepsilon_\parallel} = k_0^2 \qquad (3)$

Here, $k_0 = \omega/c = 2\pi/\lambda$ represents wavenumber in the free space, while $k_x, k_y$, and $k_z$ are the wavevector components in the media. Since in the optical range of frequencies $\varepsilon_1 < 0$, and $\varepsilon_2 > 0$, it is clear that both $\varepsilon_\perp$ and $\varepsilon_\parallel$ can change sign depending on layer thicknesses $d_1$, and $d_2$. In the $k$-space, equation (2) represents a sphere, while equation (3) represents an ellipsoid provided both $\varepsilon_\perp$, and $\varepsilon_\parallel$ are positive. For the purpose of the present paper we will confine ourselves to that case, only. We would like to emphasize here, that implementation of metallic layers is necessary to obtain positive birefringence. It is not difficult to see that in dielectric-dielectric superlattices birefringence is always negative, i.e. $\varepsilon_\perp > \varepsilon_{||}$ [4].

In fact, the exact dispersion relations are obtained via the transfer-matrix method [5]

$$\cos(k_z d) = \cos(k_1 d_1)\cos(k_2 d_2) - \frac{1}{2}(\alpha_{s,p} + \frac{1}{\alpha_{s,p}})\sin(k_1 d_1)\sin(k_2 d_2) \quad . \quad (4)$$

Here: $k_{1,2} = \sqrt{k_0^2 \varepsilon_{1,2} - k_y^2 - k_x^2}$; $\alpha_s = k_1/k_2$ for the TE-polarization, and $\alpha_p = k_1 \varepsilon_2 / k_2 \varepsilon_1$ for the TM-polarization. Equations (2, 3) are obtained from eq. (4) by simple Taylor expansion by assuming $k_1 d_1 \ll 1$; $k_2 d_2 \ll 1$, as well as $k_z d \ll 1$. As can be easily seen, eqs. (4) are periodic in $z$, while EM eqs. (2, 3) are not. To avoid this, $k_z$ in (2, 3) should be replaced by $(2/d)\sin(k_z d/2)$ to obtain quasi-effective medium (QEM) approximation [6].

In order to get insight into validity of both EM and QEM theory we have solved numerically eqs. (4), and compared the results with the corresponding approximations. Supposing $k_x = 0$, without loss of generality, we present some results in Fig. 1.

As can be seen, EM theory can be used in a limited range of parameters, and for a very limited range of $k_z$ and $k_y$. QEM is much better, but it does not reproduce the upper band in $k_y$ for the TM-polarization. The main result of this investigation is that besides the well known birefringence (circles for the TE- and ellipses for the TM-polarization), there exists the second TM-band. Thus, in a metal-dielectric superlattice, for sufficiently thin metallic layers, as well as for thin enough unit cells, we have two extraordinary TM-polarized modes and one ordinary TE-polarized mode. This novel effect can be termed tri-refringence. When one keeps a fixed metallic layer thickness, but increases the dielectric layer thickness, the ellipses become thicker and thicker, until they start to overlap (see Fig. 1, a1, b1, a2 and b2). The effect appears when $\varepsilon_\perp > \pi/k_0 d$. In that case the EM theory fails completely, while the QEM approximation follows the trend of the exact solutions, but without the second TM-band.

### 3. Surface waves at an interface of semi-infinite metal-dielectric superlattice

Now, we investigate surface waves that propagate along the boundary between semi-infinite metal-dielectric superlattice and semi-infinite isotropic dielectric, as shown in Fig. 2. In contrast to the conventional surface waves, when the superlattice is cut parallel to the layers (see e.g. [5]), we consider a metal-dielectric superlattice cut normally to the layers in contact with isotropic dielectric with dielectric permittivity $\varepsilon > 0$.

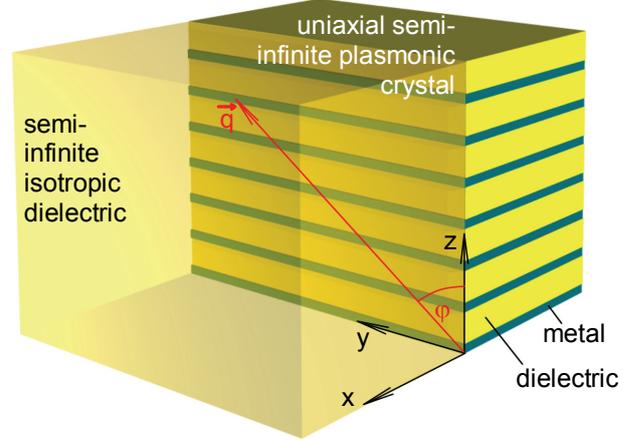

**Figure 2**. Geometry of the problem

Such configuration was investigated for the case of surface wave propagation along the z-axis [6], and it was shown that surface waves can exist if $\varepsilon_\perp < 0$, provided the use of EM theory can be justified. Here, we study the case when both $\varepsilon_\perp$ and $\varepsilon_{||}$ are positive. In order to derive a dispersion relation of the surface waves, the standard boundary conditions have to be applied. Then, the use of the EM theory leads to the following dispersion relation:

$$(k_d + k_{ex})(k_d + k_{or})(\varepsilon k_{or} + \varepsilon_\perp k_{ex}) = \\ = (\varepsilon_{||} - \varepsilon)(\varepsilon - \varepsilon_\perp) \quad (5)$$

Here, $k_d = \sqrt{q^2 - \varepsilon}$; ($k_z = k_0 q \cos\varphi; k_y = k_0 q \sin\varphi$);

$k_{or} = \sqrt{q^2 - \varepsilon_\perp}$; $k_{ex} = \sqrt{q^2 \left(\frac{\varepsilon_{||}}{\varepsilon_\perp}\cos^2\varphi + \sin^2\varphi\right) - \varepsilon_{||}}$

are the imaginary wavevector components normal to the boundary (x-direction), that have to be all positive to get solutions of the surface wave type. We see from eq. (5) that the necessary condition for the existence of surface waves is $\varepsilon_{||} > \varepsilon > \varepsilon_\perp$. These Dyakonov-like surface modes can propagate obliquely with respect to the optical z-axis, i.e. in a certain range of the angle $\varphi$; $\varphi \neq 0; \pi/2$. It is worth noting that such modes are not polaritons. They are hybrid TE-TM modes [1].

In the present paper, we confine ourselves to the study of an Ag-GaAs superlattice with $SiO_2$ cladding at $\lambda = 1.55$ μm. In that case, $\varepsilon_1 = -116 + i\,11.1$, $\varepsilon_2 = 12.4$, $\varepsilon = 2.25$. The results are presented in Fig.3. As can be seen, surface modes of this kind exist in a wide range of angles $34.4^0 < \varphi < 66.8^0$. This is a dramatic enlargement in comparison to natural anisotropic materials, and at reasonably low losses. Our analysis show that the figure of merit $\text{Re}(q)/\text{Im}(q)$ is higher than 20 in the whole range of angles where the modes exist.

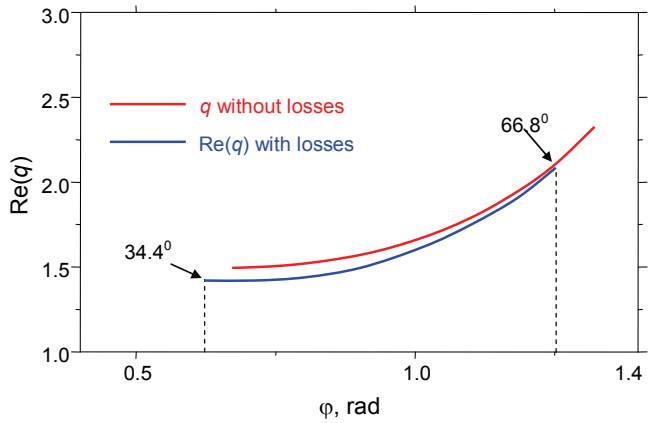

**Figure 3**. Dyakonov-like surface modes at the interface of semi-infinite plasmonic crystal (Ag-GaAs) and semi-infinite isotropic dielectric (SiO$_2$). $\varepsilon = 2.25$; $\lambda = 1.55$ μm; $\varepsilon_1 = -116 + i\,11.1$, $d_1 = 12$ nm, $\varepsilon_2 = 12.4$, $d_2 = 150$ nm.

## 4. Conclusion

We have demonstrated the existence and studied the dispersion properties of Dyakonov-like surface modes at an interface between a metal-dielectric superlattice, cut normally to the layers, and an isotropic cladding. It is shown within the framework of the effective-medium approximation that such surface waves can propagate obliquely to the optical axes, and they are hybrid TE-TM waves that exhibit dispersion and losses. However, the range of propagation angles is substantially greater than for the natural anisotropic materials at reasonably small losses.

## Acknowledgement

This work was supported by Qatar National Research Fund: NPRP 09-462-1-074. S.M.V. and Z.J. would like to acknowledge the support by the Serbian Ministry of Education and Science through the projects III 45016 and TR 32008. C.J.Z.-R. and J.J.M. wish to acknowledge the support provided by Spanish Ministry of Science and Innovation through grant TEC2009-11635.